\documentclass{aa}  
\usepackage{todonotes}
\usepackage{booktabs}
\usepackage{txfonts}
\usepackage{graphicx,amsmath,amssymb,multirow}
\usepackage{xspace}
\usepackage{color}
\usepackage{natbib}
\usepackage{nicefrac}
\usepackage{ulem}

\newcommand{\eg}{e.g.\xspace}
\newcommand{\sn}{SN\xspace}
\newcommand{\sne}{SNe\xspace}
\newcommand{\snia}{SN~Ia\xspace}
\newcommand{\sneia}{SNe~Ia\xspace}
\newcommand{\snname}[1]{SN\,#1\xspace}

\newcommand{\ie}{i.e.\xspace}

\newcommand{\myemail}{tpetr@fysik.su.se}
\begin{document}

\title{Testing for redshift evolution of Type Ia supernovae using the strongly lensed PS1-10afx at $z=1.4$}
 
 \subtitle{}

\author{T.~Petrushevska\inst{1}\fnmsep\thanks\myemail 
			\and R.~Amanullah\inst{1} \and 
			M.~Bulla\inst{1} \and
			M.~Kromer\inst{2,}\inst{3} \and
			R.~Ferretti\inst{1}\and 
			A.~Goobar\inst{1} \and
		    S.~Papadogiannakis\inst{1}
           }
\institute{Oskar Klein Centre, Department of Physics, Stockholm University, SE 106 91 Stockholm, Sweden 
	\and Zentrum für Astronomie der Universit\"{a}t Heidelberg, Institut f\"{u}r Theoretische Astrophysik, Philosophenweg 12, D-69120 Heidelberg, Germany
	\and Heidelberger Institut f\"{u}r Theoretische Studien, Schloss-Wolfsbrunnenweg 35, D-69118 Heidelberg, Germany.	
}
 
 \date{Received April 14, 2017; accepted June 8, 2017}

 \abstract
{%
  The light from distant supernovae (\sne) can be magnified through gravitational lensing when a foreground galaxy is located along the 
  line of sight. This line-up allows for detailed studies of \sne at high redshift that otherwise would not be 
  possible. Spectroscopic observations of lensed high-redshift Type Ia supernovae (\sneia) are of particular interest since they 
  can be used to test for evolution of their intrinsic properties.  The use of \sneia for probing the cosmic
  expansion history has proven to be an extremely powerful method for measuring cosmological parameters.  However, 
  if systematic redshift-dependent properties are found,  their usefulness for future surveys could be challenged. 
}
{%
  We investigate whether the spectroscopic properties of the strongly lensed and very distant \snia PS1-10afx at $z=1.4$, deviates from the well-studied populations of normal SNe Ia at nearby or intermediate distance.
}	
{%
We created median spectra from nearby and intermediate-redshift spectroscopically normal \sneia from the literature at  $-5 $ and $+1$ days from light-curve maximum. We then compared these median spectra to those of PS1-10afx.
}
{%
We do not find signs of spectral evolution in PS1-10afx.
The observed deviation between PS1-10afx and the median templates are within what is found for SNe at low and intermediate redshift. There is a noticeable broad feature centred at $\rm \lambda\sim 3500$~\AA{}, which is present only to a lesser extent in individual low- and intermediate-redshift SN~Ia spectra. From a comparison with a recently developed explosion model, we find this feature to be dominated by iron peak elements, in particular, singly ionized cobalt and chromium.
}
{}


\maketitle
\section{Introduction}
Type Ia supernovae (\sneia) have proven to be extremely powerful for studying the nature of the accelerated expansion of the Universe \citep[see \eg][for a review]{Goobar2011}.  Although Einstein's cosmological constant is not challenged by the latest results \citep{Betoule2014}, it is not unproblematic, and there is a rich choice of alternative explanations that are also consistent with the current data \citep{2017arXiv170505768D}.  These are best addressed with high-quality data at high redshift. 
At present, the sample of \sneia at $z\ge1$ is still small and mostly provided by the Hubble Space 
Telescope (HST, \citealt{2007ApJ...659...98R,2012ApJ...746...85S, 2012ApJ...746....5R, 2013ApJ...763...35R}). This will improve with future space-based surveys such as the Wide Field Infrared Survey Telescope (WFIRST; \citealt{2015arXiv150303757S}) and possibly EUCLID \citep{2014A&A...572A..80A}.
For example, WFIRST is expected to detect $\sim 2700$ \sneia up to $z\approx1.7$ and will be limited by systematic errors \citep{2015arXiv150303757S}. One important systematic effect is the possibility that the mean \sn properties evolve with redshift in a way that cannot be corrected with the current standardization techniques  \citep{2007ApJ...667L..37H, 2008ApJ...684...68F, 2010MNRAS.406..782S}. 

Properties of galaxies such as mass, age, dust and metallicity change with cosmic time, and this may also affect the \sne that they host.  In order to test for systematic variations of the \snia population with time, mean spectra at low, intermediate and high redshift have been constructed to measure the possible evolution of \snia spectral properties \citep{Ellis2008,Sullivan2009,Foley2012,Maguire2012}.
These studies did not find any change in the optical part, while the near-UV showed increased scatter. In particular, \citet{Maguire2012} detected a $\sim3\sigma$  excess of the pseudo-continuum flux for the intermediate-redshift mean spectrum compared to the nearby one.   \citet{Sullivan2009} found a slight decrease with redshift in the strength of intermediate-mass element features (Si\,{\sc ii}, Ca\,{\sc ii}, and Mg\,{\sc ii}), but concluded that this is not an evolutionary effect of intrinsic \sn properties.

While the large diversity of \sneia in the UV domain is supported by some theoretical studies that predict sensitivity to differences in explosion mechanisms, progenitor age, metallicity and environment, it is not certain which of these play the dominant role \citep[see \eg][]{2014Ap&SS.351....1P}. For instance,  \citet{2003ApJ...590L..83T} studied  the production of $^{56}\rm Ni$ in relation to the initial metallicity of the progenitor white dwarf. 
	The reasoning behind this is that the presence of metals in the white dwarf translates into lower proton-to-nucleon ratios, which leads to the production of more neutron-rich nuclei of the iron group elements (IGEs) and less radioactive  $^{56}$Ni \citep{1998ApJ...495..617H}. Since UV spectra are characterized by strong line-blanketing from IGE lines,  higher (lower) UV fluxes are typically expected for lower (higher) progenitor metallicities \citep{2000ApJ...530..966L}. However, despite these efforts, there is no consensus yet regarding the relationship between progenitor metallicity and  SN spectra.

At present, the signal-to-noise ratio (S/N) and the resolution of the high-$z$ sample spectra from HST \citep{2007ApJ...659...98R} is not good enough for a detailed study. Thus, even individual high S/N events at high redshift could be of interest if they show discrepancies from what has been observed in the large low-$z$ samples. This unique opportunity arose with the discovery of the strongly lensed SN Ia PS1-10afx. Here, we analyse this first high-S/N rest-frame near-UV SN Ia spectrum. Another magnified \snia dubbed \snname{HFF14Tom}at $z=1.3457$ was found behind a galaxy cluster \citep{2015ApJ...811...70R}, but because of the modest magnification, it had a lower S/N and is therefore not considered in this work. 

 The source PS1-10afx was discovered on 2010 August 31.35 UT with the 1.8 m Pan-STARSS1 (PS1) telescope as part of the PS1 Medium Deep Survey. Follow-up observations and near-infrared (NIR) spectroscopy revealed a redshift of $z=1.3883$. Based on the redshift, the transient was brighter than any previously observed SN. Therefore,  \citet{2013ApJ...767..162C} (C13) classified PS1-10afx as a new type of superluminous~\sn, but the shape of the light curve did not fit those expected for superluminous~\sne well.  Later,  \citet{2013ApJ...768L..20Q} classified PS1-10afx as an \snia magnified $\sim30$ times. This magnification was caused by a galaxy lens whose redshift was determined from the Mg\,{\sc ii} and Ca\,{\sc ii} in the Keck/LRIS spectrum taken three years after the \sn explosion (\citealt{2014Sci...344..396Q}, Q14).

\section{Analysis}







In this work we make use of three of the spectra presented in C13 obtained with ground-based facilities. There are two GMOS-S and GMOS-N optical (observer frame) spectra, and one  FIRE/Magellan NIR observer frame spectrum.
 The optical spectra are at phase of $-5.0$ and $+1.2 \;\rm d$ with respect to the B-band maximum rest-frame epoch, while the NIR  spectrum is from $-2.1 \;\rm d$\footnote{There is also one NIR spectrum at phase $+2.0 \; \rm d$, but because of the low S/N ratio and limited wavelength coverage it is not used in this study.}. 
The PS1-10afx spectra have been corrected for Galactic extinction $\rm E(B-V)=0.05$~mag \citep{2011ApJ...737..103S}. Since the observed SN colours suggest no extinction (Q14), no de-reddening to correct for host extinction was applied. We did, however, detect interstellar Mg\,{\sc ii} absorption at $2796$~\AA{} and $2804$~\AA{} at the host galaxy redshift, which according to the empirical relation found by  \citet{2008MNRAS.385.1053M} would indicate
non-negligible extinction. However, as shown in \citet{2015MNRAS.453.3300A}, reddening of \sneia correlates very poorly with Mg\,{\sc ii} absorption.











The Si\,{\sc ii}~$6355\:\AA$ feature is typically used to classify \sneia.
As discussed in \citet{2013ApJ...768L..20Q}, this feature is not prominent in the optical rest-frame spectrum, although this is uncertain because of the low S/N ratio in the red part of the spectrum (see also Section~\ref{sec:disc}). There are peculiar \snia subtypes where this feature is weak (1999aa-like, see e.g. \citealt{2004AJ....128..387G}) or even almost completely absent (1991T-like SNe~Ia, see \citealt{1992ApJ...384L..15F}).  However, 1999aa- and 1991T-like \sneia have slower decline rates with  $\Delta m_{15}(B ) \lesssim 0.9$~mag,  in contrast with PS1-10afx that has $\Delta m_{15}(B )=1.22\pm0.09$.  Moreover, they are known to have bluer UV colour evolution than normal \sneia \citep{2014ApJ...787...29B,Smitka2015}. The comparison of the light-curve shapes and UV rest-frame colours predicted from the \snia templates in \citet{2013ApJ...768L..20Q} further confirms PS1-10afx  as a normal SN~Ia.

\subsection{Velocity and pEW of the Ca~{\sc ii}~H\&K feature}
First, we focus on the prominent \sn absorption feature at $\sim 3700$~\AA{}, to which Ca\,{\sc ii}~H\&K,  Si\,{\sc ii} and
various ionization states of the iron-peak elements contribute significantly. The Ca\,{\sc ii}~H\&K feature strength, expressed as pseudo-equivalent width (pEW), and the line blue-shift with respect to the rest wavelength of $3945.12$~\AA{}, expressed as expansion velocity, have previously been used as a diagnostic for \snia evolution.  \citet{2007A&A...470..411G}  measured Ca\,{\sc ii}~H\&K expansion velocities and pEWs for 13 \sneia at $0.279<z<0.912$, and concluded that they did not deviate significantly from the nearby \sneia. 

 We measured the expansion velocities and pEWs for PS1-10afx and the results are shown in Table~\ref{table:calcium}. The large uncertainties in the velocity measurements are due to the overlapping [O\,{\sc ii}] emission line.  
The given errors  represent the two extreme values of the velocity when we assume that the minimum of the Ca\,{\sc ii}~H\&K feature is at either side of the host galaxy emission line.  We measure an expansion velocity of $\sim 16500$~km/s, consistent with the value previously reported by C13, and a pEW at maximum of $\sim 80$~\AA{}. The mean value of the Ca\,{\sc ii}~H\&K  velocity in the \citet{2013ApJ...773...53F} nearby sample is  $\sim 15000\pm3000$~km/s and pEW of $\sim120\pm30\AA$, measured around B-band maximum brightness. With their large uncertainties, the values for PS1-10afx are statistically consistent with the nearby sample.

\begin{table}
	\caption{PS1-10afx Ca\,{\sc ii}~H\&K feature properties.}\label{table:calcium} \centering \begin{tabular}{l cc} \hline\hline Spectrum  & $v_{\rm Ca\:II}$  & pEW$_{\rm Ca\,{\sc II}}$ \\
			phase	&  ($10^3$ km/s) & $ (\AA)$ \\
		\hline
	
		-5.0 d &  $16.5\, (2.8) $  & $88\, (8)\: $ \\
		-2.1 d& $16.7 \, ( 3.0) $  & $103\,  ( 13)\: $ \\
		+1.2 d & $16.0 \, (3.0) $ & $75\,  ( 19)\: $ \\
		\hline

	\end{tabular} 
			\tablefoot{Uncertainties are given in parentheses.}
\end{table}

\subsection{Spectral comparisons}\label{sec:nearby}
\begin{figure*}[htbp]
	\begin{center}
		\includegraphics[width=\hsize]{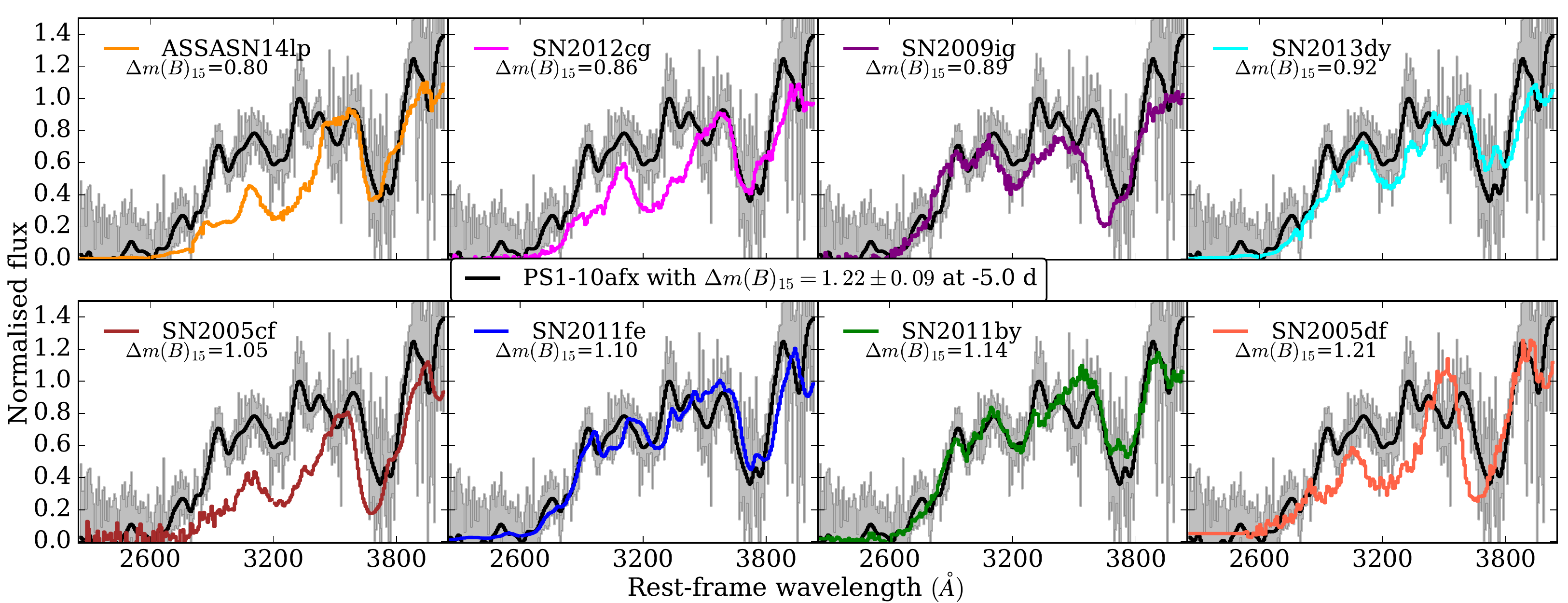}
		\caption{PS1-10afx at phase $-5.0$ d (grey line) and nearby \snia spectra at similar phases. When there were two spectra close to the required phase, we took the average. The black line shows the PS1-10afx spectrum rebinned by a factor of 10. The spectra were normalised at $\sim 4000$~\AA{}. The nearby spectra are ordered according to their light-curve shape parameter $\Delta m_{15}(B)$. }
		\label{nearby}
	\end{center}
\end{figure*}

Space-based facilities such as HST and the Swift satellite, have allowed studies of nearby SNe~Ia in the UV \citep[see \eg][]{2014ApJ...787...29B}. We can use the spectroscopically normal SNe~Ia from this sample at similar phase with PS1-10afx to make a comparison.  As SN~Ia spectra evolve rapidly, particularly in the UV \citep[see e.g.][]{Hook2005}, we restrict the epochs of the nearby UV spectra to phases within $\pm2.5\;\rm d$ from the PS1-10afx epochs. First, we use the optical spectrum at $-5.0 \;\rm d$, since it has the highest S/N.  At phases $-7.5$ to $-2.5\; \rm d$, we found eight \sne from the literature and their properties are shown in Table~\ref{table:nearby}.  \snname{2014J} also meets the criteria, but since it is highly reddened by its host galaxy \citep[\eg][]{2014ApJ...784L..12G, 2014ApJ...788L..21A}, it was not included in this analysis. We used the Swift spectra that have been re-reduced by \citet{2016PASP..128c4501S} with their novel decontamination technique. The spectra were corrected for Galactic and host extinction. We used the parameterization of Milky Way extinction from \citet{F99} and $R_V=3.1$. Following previous studies with SN~Ia UV spectra  \citep{2016MNRAS.461.1308F}, the spectra were normalised with a top-hat passband at $\lambda \sim 4000 \AA$. When two spectra were close to the required phase, we took the average  to give a mean phase of $-5.0$~d. 
In Figure~\ref{nearby} we show the comparison between the early PS1-10afx spectrum  and the low-$z$ \snia sample. 

 There is no perfect agreement  to PS1-10afx with any of the observed nearby objects. However, the shape of the continuum and most of the features agree well with those of the spectra of \snname{2011fe}, 2011by and 2013dy.  In contrast, the blue component of the Ca\,{\sc ii}~H\&K feature in \snname{2009ig},  \snname{2005cf} and \snname{2005df} differs from that of PS1-10afx.  It has been proposed that the blue component is caused by either high-velocity calcium or  Si\,{\sc ii}~3859\,\AA{}  \citep[][and references therein]{2013MNRAS.435..273F}.

In Figure~\ref{nearby} we order the nearby SN spectra according to their light-curve shape parameter $\Delta m_{15}$, motivated by \citet{2008ApJ...684...68F,2016MNRAS.461.1308F}, who argued that  the UV spectral continuum is driven primarily by this parameter. In particular, \citet{2016MNRAS.461.1308F}, after normalising at $\sim 4000$~\AA{}, found that SNe with lower values of  $\Delta m_{15}$ have a higher flux level at $\sim 3000$~\AA{}. However, in the limited range of $\Delta m_{15}$ values from 0.8 to 1.2, which our sample spans, this correlation is not obvious. However, we do find that the flux at $\sim 3025$~\AA{} is between $\sim35$ and $65\%$ of that at $\sim 4000$~\AA{}. Although our spectra are at earlier epochs than those of \citet{2016MNRAS.461.1308F}, the flux ratio that we measure is consistent with their findings (upper left panel of their Figure~3). 

\begin{figure}[htbp]
 	\begin{center}
 		\includegraphics[width=\columnwidth]{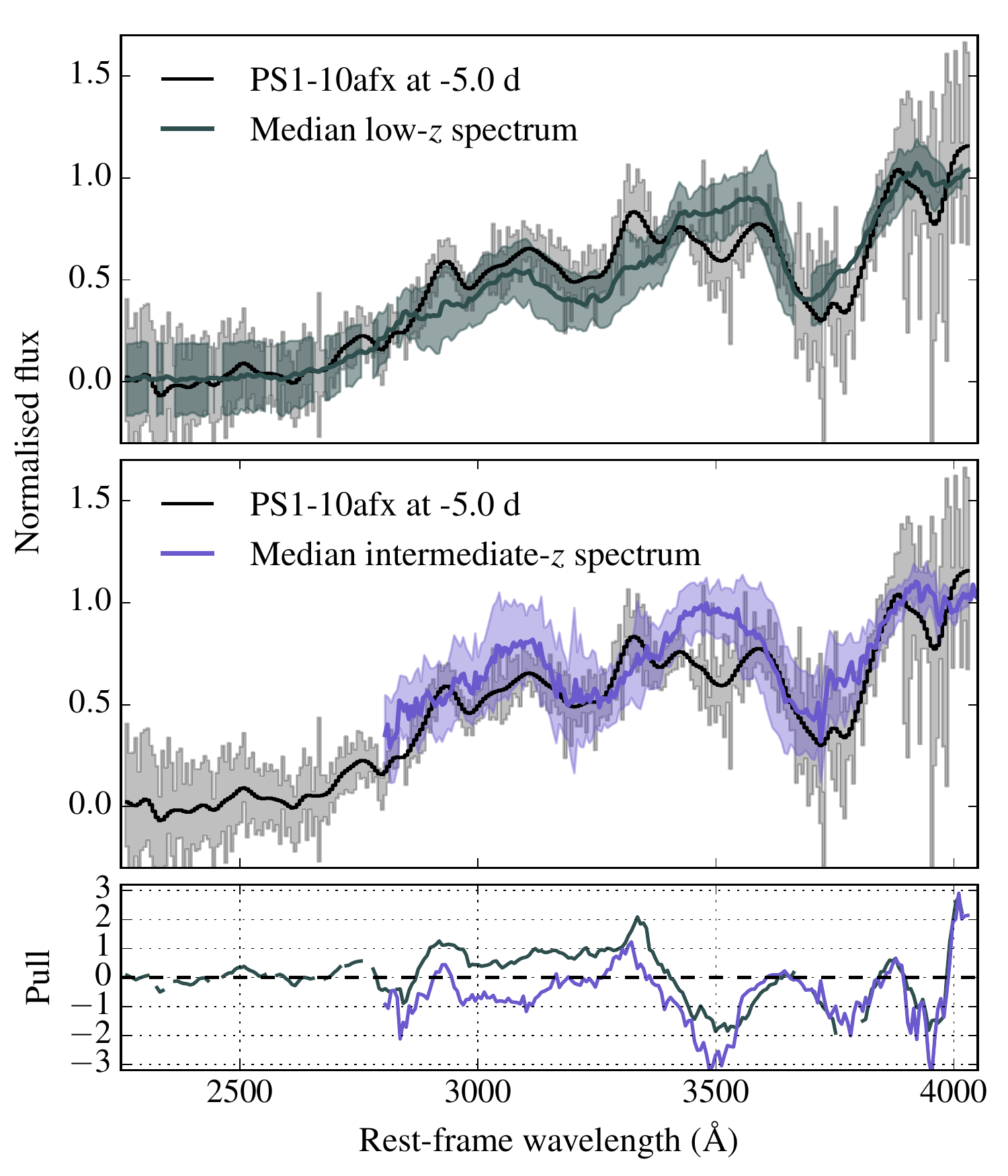}
 		\caption{PS1-10afx at $-5.0$ d  compared with low- and intermediate-redshift median spectra constructed from normal \sneia at similar phases. The spectra were normalised at $\sim 4000$~\AA{}. The shaded regions represent the root mean square of the sample in green and blue for the nearby and intermediate \sneia, respectively.
 			 The lower panel shows the pull for the low- (green) and intermediate-redshift (blue) spectrum.  }
 		\label{early}
 	\end{center}
 \end{figure}

\begin{figure}[htbp]
	\begin{center}
		\includegraphics[width=\columnwidth]{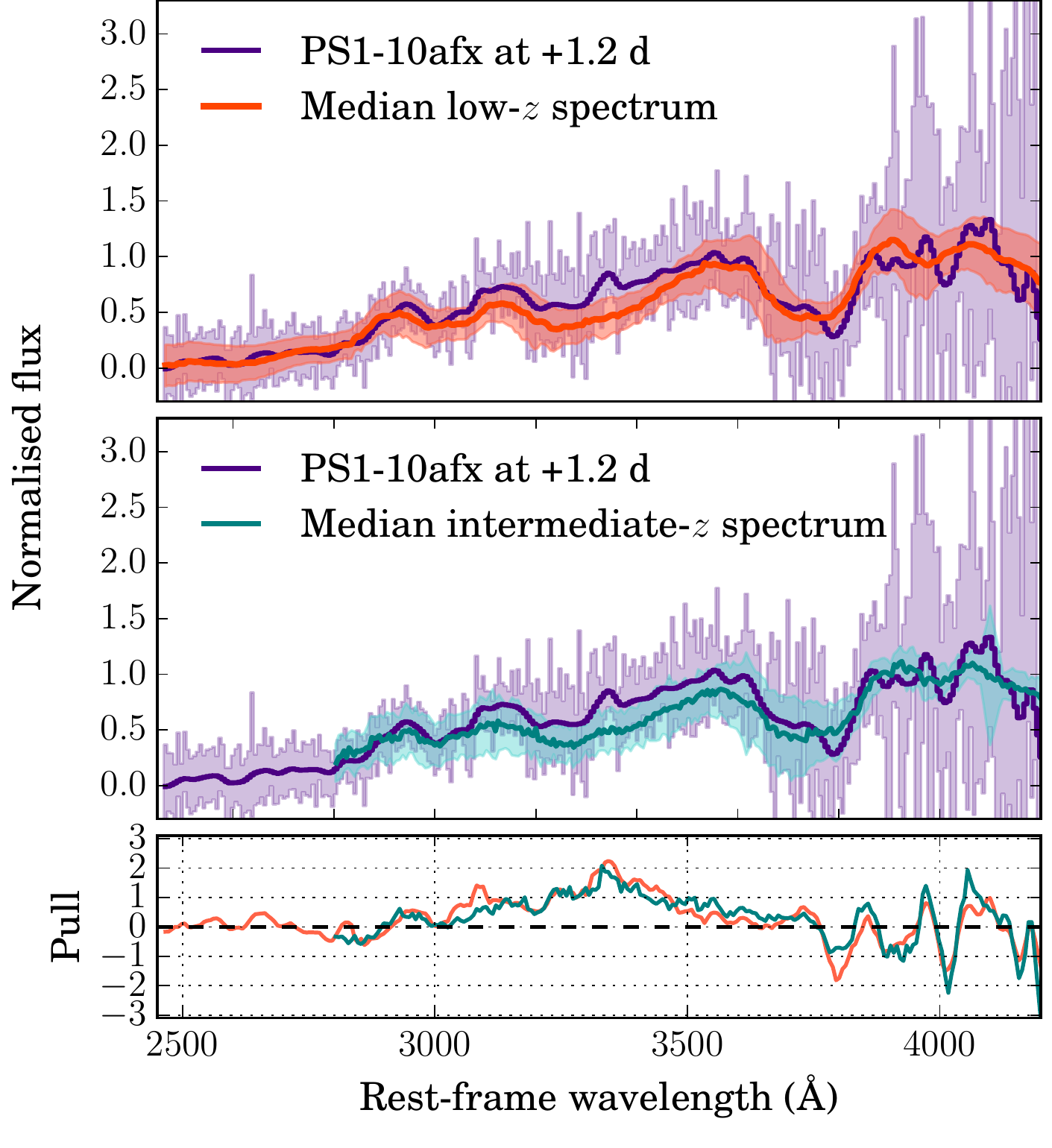}
		\caption{PS1-10afx at $+1.2$ d  compared with low- and intermediate-redshift median spectra constructed from normal \sneia at similar phases. The shaded regions represent the root mean square of the sample in pink and cyan for the nearby and intermediate \sneia, respectively.
			The lower panel shows the pull for the low- (pink) and intermediate-redshift (cyan) spectrum. }
		\label{max}
		
	\end{center}
\end{figure}

\begin{figure}[htbp]
	\begin{center}
		\includegraphics[width=\columnwidth]{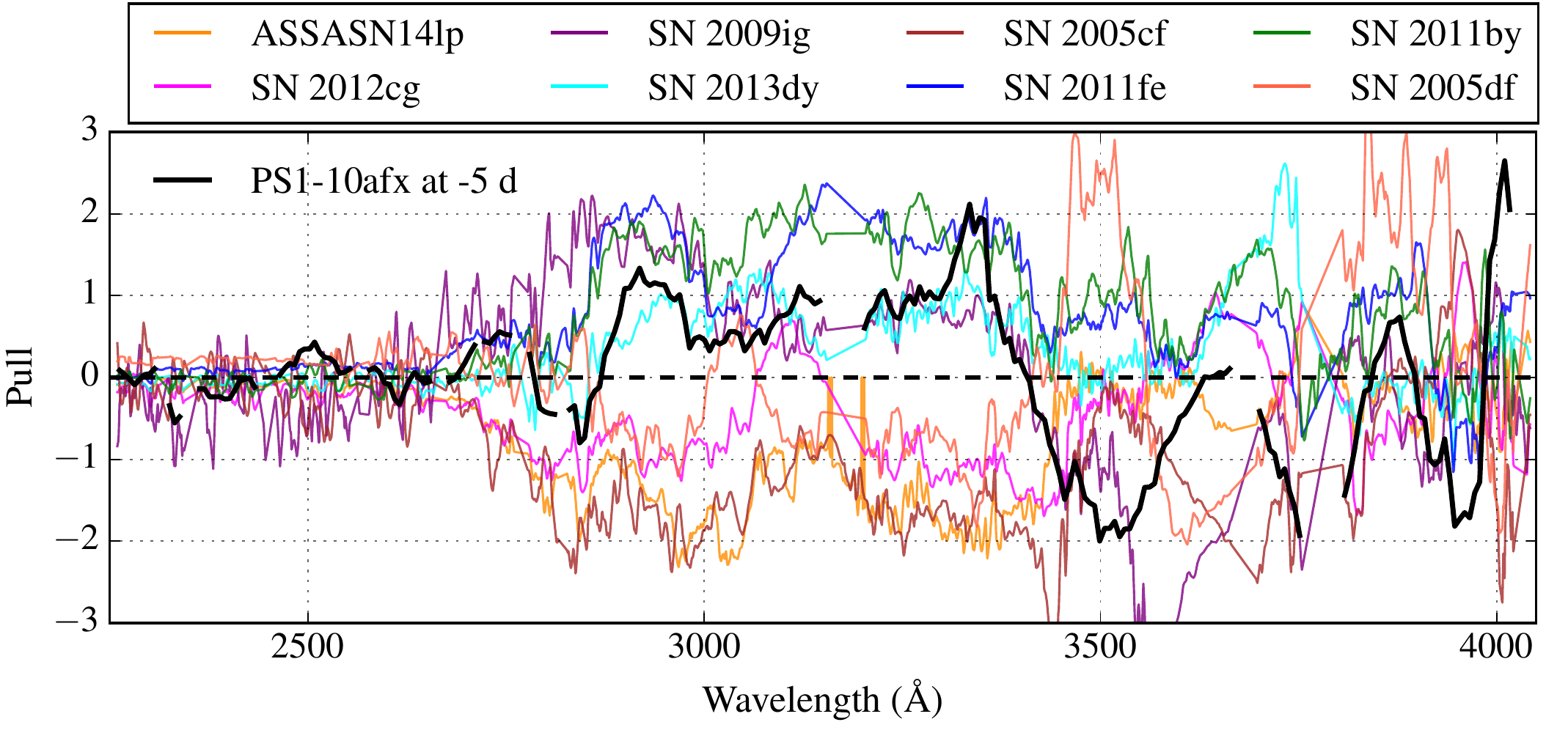}
		\caption{Pull of PS1-10afx at $-5.0\: \rm d$ compared with the low-redshift median spectrum constructed from all the \sneia in Table \ref{table:nearby}, shown by the black line. The coloured lines represent the pulls of the individual nearby \sneia compared to the median spectrum constructed by leaving that one out from all the \sneia spectra.  }
		\label{jackknife}
	\end{center}
\end{figure}

Using published data of normal nearby SNe~Ia, we also constructed median spectra at phases similar to those of PS1-10afx.  When building the median spectrum, we used the same restriction on phase and normalisation as outlined above. At phases close to $-5\; \rm d$, we used the SNe~Ia in Table~\ref{table:nearby} and shown in Figure~\ref{nearby}, while at phases close to $+1.2\; \rm d$, we also found eight \sne that are in the appropriate range, and we list them in Table~\ref{table:max}. In the upper panels of Figure~\ref{early} and \ref{max} we compare the median spectra with the rms scatter at low redshift with spectra of PS1-10afx at $-5\; \rm d$ and $+1.2 \; \rm d$, respectively. To illustrate how much PS1-10afx differs from the median spectra, we also plot the pull values (\ie the residuals divided by the square root of rms squared) in the lowest panel. In order to assess the significance of the observed deviations, we also applied a jack-knife approach for the nearby sample. The median spectrum was constructed after leaving out one of nearby spectra at the time.  This spectrum is then compared to the median spectrum in the same manner as for PS1-10afx.  The result is shown in Figure  \ref{jackknife}, from which we can conclude that the deviations of PS1-10afx are similar to the results from the jack-knife exercise, and all pulls are within 3 at all wavelengths.

\begin{table*}
	\caption{Properties of the nearby \sneia used for the early-time (-5 d) median spectrum.}\label{table:nearby} \centering \begin{tabular}{ccccccccc} \toprule 
		SN name & ASASSN-14lp &\snname{2012cg} & \snname{2009ig}& \snname{2013dy} &\snname{2005cf} & \snname{2011fe} & \snname{2011by} & \snname{2005df} \\
		
		z$_{\rm helio}$ &0.005101$^a$ & 0.001447$^c$ & 0.007589$^c$ & 0.00383$^g$ & 0.006$^h$ & 0.002000$^j$ & 0.002843$^c$ &0.004350$^n$\\
		$A_V$ & $0.99$$^a$  & 0.34$^d$ & 0.01$^f$  & 0.64$^g$ & 0.10$^h$ & $0.08$$^j$&$0.03^l$&0.001$^n$\\
		MW $A_V$ & $0.0682$  & 0.0558$^d$  & 0.088$^f$ & 0.46$^g$ & 0.269$^h$ & 0.024$^j$&$0.038^l$&0.090$^n$\\
		$\Delta m_{15}(B)$ & $0.80(0.05)$$^a$&$0.86(0.02)$$^d$ &  $0.89 (0.02)$$^f$  & $0.92 (0.03)$$^g$  &$1.05(0.03)$$^h$ &$1.10(0.04)$$^j$ & $1.14(0.03)$$^m$ &$1.21(0.04)$$^n$\\
		instr. &STIS/HST$^b$ & SWIFT$^e$ & SWIFT$^e$  & STIS/HST$^g$   & SWIFT$^i$ & STIS/HST$^k$&SWIFT$^e$&SWIFT$^e$\\
		sp. phase & $-4.4$ d$^b$  &$-7.2$ d$^e$&$-4.2$ d$^e$ & $-6.6$ $-2.5$ d$^g$ & $-5.8$ d$^i$ &$-6.9$ $ -2.9$ d$^j$ &$-7.9$ $-3.9$ d$^e$ &$-5.2$ d$^e$ \\ 
		\bottomrule
	\end{tabular} 
	\tablefoot{Uncertainties of the light-curve shape parameter $\Delta m_{15}(B)$ are given in parentheses.\\ References: $^a$\citet{Shappee16}, $^b$ \citet{2016MNRAS.461.1308F}, $^c$From NED,  $^d$\citet{2016ApJ...820...92M}, $^e$\citet{2016PASP..128c4501S}, $^f$\citet{2012ApJ...744...38F}, 	$^g$ \citet{Pan2015}, $^h$\citet{2009ApJ...697..380W}, $^i$\citet{2009ApJ...700.1456B},   $^j$\citet{2013A&A...554A..27P}, $^k$\citet{2014MNRAS.439.1959M}, $^l$\citet{2015MNRAS.446.2073G}, $^m$\citet{2013MNRAS.430.1030S}, $^n$\citet{2010ApJ...721.1627M}.
	}
\end{table*}

\begin{table*}
	\caption{Properties of the nearby \sneia used for the median spectrum at phase $+1.2 \: \rm d$.}\label{table:max} \centering \begin{tabular}{ccccccccc} \toprule
		SN name & \snname{2009ig} & \snname{2013dy} & \snname{1981B}&\snname{2001ba}& \snname{ 2011fe} &\snname{2011by}  &\snname{ 2005df}&\snname{1992A}  \\
		z$_{\rm helio}$ & 0.007589 & 0.00383 & 0.005101$^c$& 0.029557$^c$ &0.002000& 0.002843& 0.004350 &0.001447$^c$\\
		$A_V$ &  0.01  & 0.15 & 0.35$^d$ & 0.016$^d$  & 0.08 & 0.03&0.001&0.016$^d$\\
		MW $A_V$& 0.088 & 0.92 & 0.0682$^d$ & 0.176$^d$ & 0.024 & 0.038 & 0.090&0.046$^d$\\
		$\Delta m_{15}(B)$ & $0.89(0.02)$ &$0.92(0.03)$ & $1.01(0.10)$$^d$ &  $1.01(0.11)$$^d$  & $1.10(0.04)$&$1.14(0.03)$ & $1.21(0.04)$ &$1.47(0.11)$$^d$\\
		instr. & SWIFT$^a$ & STIS/HST$^b$ & IUE$^d$   & STIS/HST$^d$  & SWIFT$^a$ & SWIFT$^a$ & SWIFT$^a$ & IUE$^d$\\
		sp. phase &+1.6$^a$ & $+1.2$ d$^b$ & +1.8 d$^d$ & +3.8 d$^d$  & $+0.1$ $+3.0$ d$^a$  &$+1.1$ d$^a$  & $+1.6$ d$^a$& $-0.2$ $+2.4$ d$^d$ \\ 
		\bottomrule
	\end{tabular} 
	\tablefoot{Uncertainties of the light-curve shape parameter $\Delta m_{15}(B)$ are given in parentheses.\\ References: $^a$\citet{2016PASP..128c4501S}, $^b$\citet{Pan2015}, $^c$From NED, $^d$\citet{2008ApJ...686..117F}.
	}
\end{table*}

We also compared PS1-10afx with median spectra at intermediate redshift.  \citet{Ellis2008} used the LRIS/Keck to observe a homogeneous sample of 36 SNe Ia at 0.1< z <0.8 discovered with the Canada-France-Hawaii Telescope Supernova Legacy Survey (\citet{2006A&A...447...31A}; SNLS). From these host-subtracted spectra, we constructed one median spectrum at phase  $-5.0\: \rm d$  and another at  $+1.2\: \rm d$ , with the same requirement for the spectra phases as for low redshift. For the early median spectrum, 9 spectra satisfy the requirement and their mean redshift is $z_\textrm{mean}=0.44$. 
For the median spectrum at $+1.2\: \rm d$,  11 spectra  satisfy the requirement and their mean redshift is $z_\textrm{ mean}=0.46$. The comparison of the early and maximum PS1-10afx spectra to the median intermediate-$z$ spectra is shown in the middle panels in Figure~\ref{early} and \ref{max}. The intermediate-redshift templates look similar to their low-redshift counterparts, with some discrepancies at  $\lambda <3000$~\AA{},  consistent with previous results \citep{Sullivan2009,Maguire2012}.

The agreement of the PS1-10afx spectra with the \snia templates is good, within the range of variability of the nearby and intermediate sample which is less than $3 \sigma$.  We discuss the largest discrepancy that we find in the following section.

 \section{Discussion}\label{sec:disc}

  The PS1-10afx spectra are statistically consistent with the diversity seen in nearby and intermediate-redshift observations at comparable epochs, despite the modest size of the samples.

We note that the  early PS1-10afx spectrum shows an extended feature in the region centred at $ \lambda \sim 3500\:\AA$$ $. This feature is also observed in SN~Ia spectra and is present to a lesser extent in some of the individual spectra that  make up the median low- and intermediate-redshift spectra, for instance in \snname{2013dy} (see Figure~\ref{nearby}). 
The lensed \snia, \snname{HFF14Tom} at $z=1.3457$, can  be matched with nearby normal \sneia \citep{2015ApJ...811...70R} and is broadly consistent with PS1-10afx. However, \snname{HFF14Tom} did not exhibit a similar feature at $ \lambda \sim 3500\:\AA$$ $, although this is uncertain as it could have been lost in the low S/N. The difference between the spectra of PS1-10afx and SN~2011fe over the range of $3300$--$3700\;\AA$$ $ 
appears to have an $\rm EW\sim40$~\AA{}. The interstellar Ca\,{\sc ii}~H\&K absorption lines of the lensing galaxy fall in this region, as reported by Q14. However, the interstellar medium (ISM) lines can have an EW of at most a few \AA{} and only contribute a small amount ($<10$\%) to the observed difference between the spectra. Furthermore, the feature  spans $\sim 200\;\AA$ of the spectrum. Given the resolution of the spectrum, the ISM lines of the lensing galaxy will have a FWHM of $\sim10\;\AA$ and be separated by $\sim30$~\AA{} at the host galaxy rest frame.  This suggests that the profile of the feature is intrinsic to the SN itself.

To investigate the nature of this feature, we examined the near-UV spectrum predictions from a recently developed multi-dimensional explosion model. Specifically, for our comparison we selected the three-dimensional N100 model \citep{Ropke2012,Seitenzahl2013}. This model describes the thermonuclear explosion of a carbon-oxygen white dwarf at the Chandrasekhar mass limit and is found to reproduce observables of normal SNe~Ia in the optical reasonably well \citep{Ropke2012,Sim2013,2016MNRAS.462.1039B}.

\begin{figure}[htbp]
	\begin{center}
		\includegraphics[width=\columnwidth]{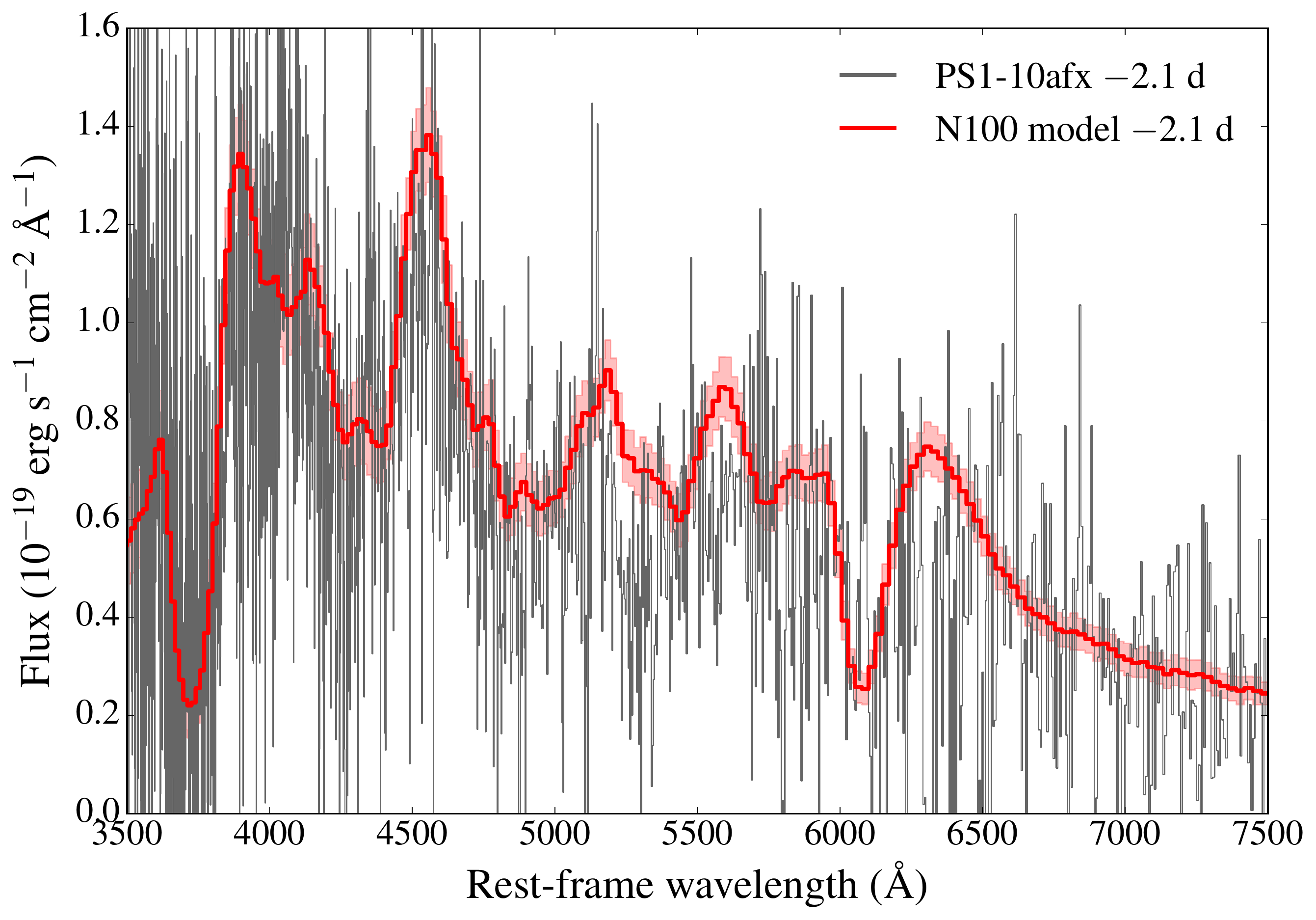}
		\caption{Comparison between the optical spectrum of PS1-10afx and the spectrum predicted by the N100 model of \citet{Seitenzahl2013}. Unlike  Figures 1--4, absolute fluxes are reported here. The observed spectrum is shown in grey and has been demagnified assuming $\mu=30.8$ from \citet{2013ApJ...768L..20Q}. An angle-averaged spectrum for the three-dimensional N100 model is shown in red, while light red regions show the dispersion that is due to viewing-angle effects.}
		\label{modelN100opt}
	\end{center}
\end{figure}

\begin{figure}[htbp]
	\begin{center}
		\includegraphics[width=\columnwidth]{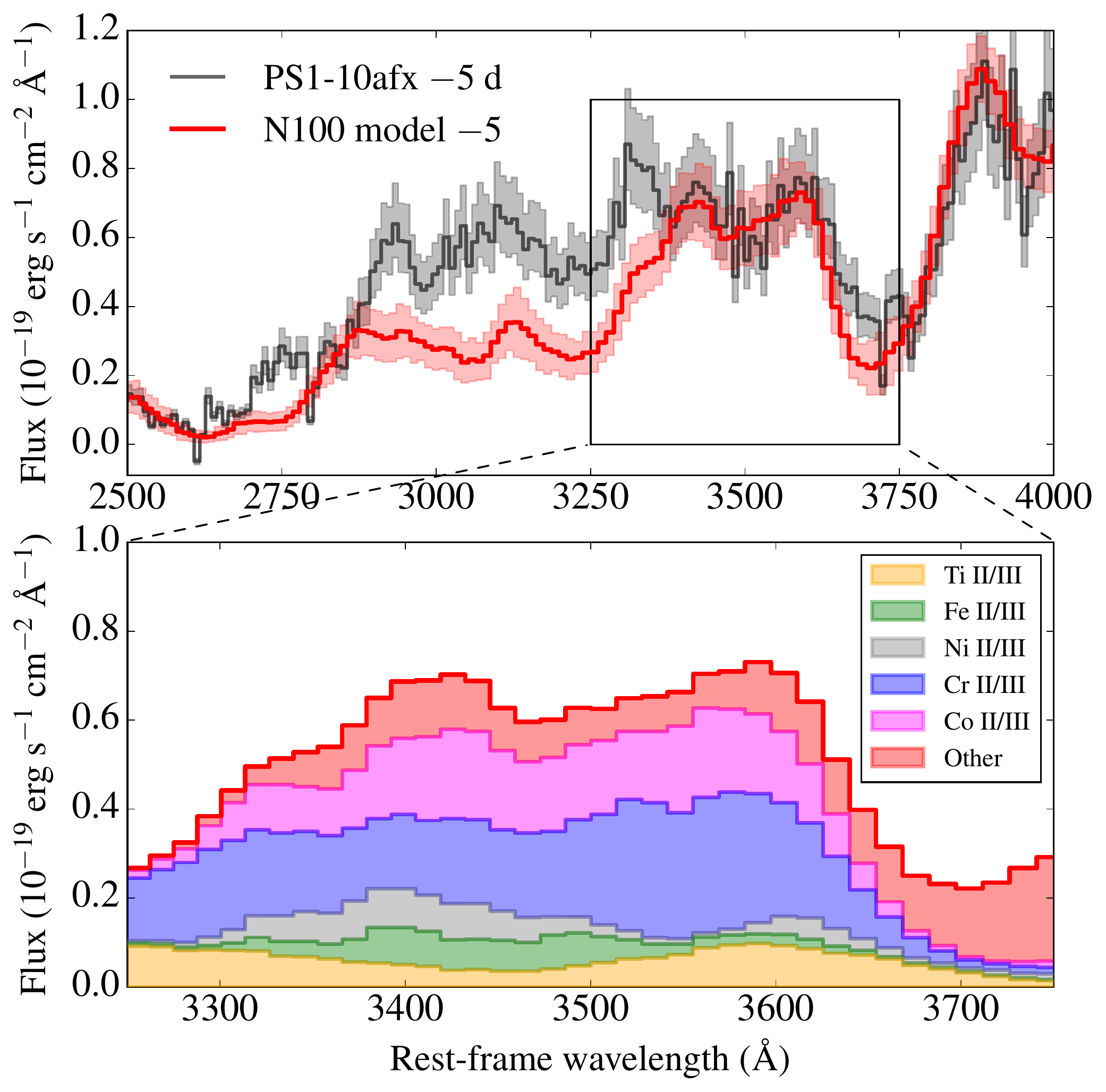}
		\caption{\textit{Upper panel.} Same as Figure~\ref{modelN100opt}, but in the near-UV. The observed spectrum was rebinned by a factor of 15 to have a similar bin size ($\sim$10~\AA) as in the model spectrum. \textit{Bottom panel.} Zoom of the N100 model into the spectral region around 3500~\AA. 
			The region below the total flux (red line) is colour-coded to indicate the relative contributions of different element transitions.}
		\label{modelN100}
	\end{center}
\end{figure}

Figure~\ref{modelN100opt} shows a comparison between the optical spectrum of PS1-10afx at $-2.1$~d relative to maximum and the spectrum predicted by the N100 model at the same epoch.  The flux spectrum of PS1-10afx has been demagnified assuming a magnification factor of $\mu=30.8^{+5.6}_{-4.8}$ \citep{2013ApJ...768L..20Q}, while the model spectrum has been scaled to the distance of PS1-10afx. Although the S/N ratio in the observations prevents us from investigating individual features in great detail, we note a good match between model and data in both luminosity and overall spectral shape. The top panel of Figure~\ref{modelN100} shows the near-UV spectrum of PS1-10afx at $-5$~d relative to maximum compared to the synthetic spectrum of the N100 model at the same epoch. The absolute fluxes are scaled as described above. We note a reasonable match between synthetic and observed spectra (but see below) not only in terms of the overall spectral shape but also across individual spectral features (see e.g. the Ca\,{\sc ii}~H\&K line). In particular, the N100 model predicts a feature in the wavelength region around $3500~\AA$, with both line strength and width comparable to those observed in PS1-10afx. As shown in the bottom panel of Figure~\ref{modelN100}, the spectral region between $\sim3300$ and $3600$~\AA{} is dominated by transitions from iron-group elements. Specifically, the characteristic feature around 3500~\AA{} appears to originate mostly from singly and doubly ionized Co and Cr transitions. 

We note that the model predicts flux at wavelengths shorter than $3400$~\AA{} that is lower than the early near-UV PS1-10afx spectrum. We caution that absolute comparison studies for lensed \sne are not straightforward as their absolute flux depends on the exact magnification factor assumed. Nevertheless, one intriguing possibility is that the flux difference observed in the near-UV might be due in part to metallicity effects. The initial carbon-oxygen white dwarf in the N100 model of \citet{Seitenzahl2013} contains 2.5\% of $^{22}\rm Ne$ to mimic the effects of a solar metallicity zero-age main-sequence  progenitor star on the nuclear burning. Thus, it is tempting to think that lower progenitor metallicities might result in higher UV fluxes and thus provide better agreement with PS1-10afx. Q14 found the best-fit value of PS1-10afx host to be $\sim 0.4$ times the solar metallicity, supporting this possibility. We note, however, that \citet{Seitenzahl2013} did not include IGEs in solar composition in their progenitor models, which could significantly affect the predicted UV spectra.

 Although \citet{Seitenzahl2013} computed alternative versions of the N100 models with different metallicities, no radiative transfer calculations have been performed for these models and it is thus unclear to which extent metallicity would affect the UV and optical part of the spectrum.   In an earlier work using the one-dimensional W7 model \citep{1984ApJ...286..644N,1986A&A...158...17T},  \citet{2000ApJ...530..966L} studied the impact of a wide range of progenitor metallicities on the observed spectra.  In their Figure~9, the $3500$~\AA{} feature is also visible and its strength depends on the assumed metallicity, although there is no simple linear relation between the feature strength and the metallicity.     In addition, metallicities of well-studied nearby \snia host galaxies show a great range of variation \citep{2012ApJ...750..120Z}, thus the metallicity measured for PS1-10afx host is not exceptional. In conclusion, PS1-10afx does not show signs of evolution, and this is in agreement with the sample of nearby normal SNe Ia.

\section{Summary and conclusions}
 We have compared the spectra of a magnified high-redshift \snia at $z=1.4$ and  spectra compiled from nearby and intermediate-redshift normal \sneia. Comparing rest-frame wavelengths from 2500 to 4000~\AA{}, we found PS1-10afx to be in good ($< 3 \sigma$) agreement with the nearby sample. We used the spectral time series to measure the  pEW and velocity and found them to be statistically consistent with those for nearby observations. From this comparison, there is no statistically significant evidence that this  \snia exhibits a redshift evolution. The spectrum of PS1-10afx at $-5.0$~d shows a broad feature at $\lambda \sim 3500 \:\AA{}$, which is only mildly present in nearby and intermediate-redshift \sneia. Because this feature is dominated by IGEs, it seems possible that metallicity effects could play a role in this difference. However, with the current uncertainties of both models and data, it is not possible to place any firm constraints.

In the second half of the 2020s, the wide-field space-based SN survey, WFIRST, will provide numerous \snia light curves at high-$z$ to be used for cosmological studies, thus systematic uncertainties are expected to limit the accuracy in the measurements of cosmological parameters. The possible evolution of \snia properties with redshift could contribute significantly to the systematic error budget. With PS1-10afx,  we were able to test this possibility several years before planned \sn surveys come online. 
As the study done here is limited to one \snia, it is quite possible that not all \sneia at $z\sim1.4$ are represented by PS1-10afx, thus that all high-$z$ are similar to the normal nearby \sneia. Furthermore, we emphasize that even though PS1-10afx is consistent with low- and intermediate-$z$ data, this does not prove that the average SED does not evolve with redshift. However, spectroscopically normal \sneia must be present in the high-redshift Universe, so the effort of the future surveys consist of  selecting these events as early as possible from the light curves in time to ask for a follow-up spectrum. 

Our study was made possible thanks to the magnification from the foreground lens galaxy, otherwise PS1-10afx would have been undetected. Wide-field ground-based surveys will see first light very soon, the Zwicky Transient Facility in 2018 (ZTF; \citealt{2016AAS...22731401K}) and the Large Synoptic Survey Telescope in 2021 (LSST; \citealt{2009arXiv0912.0201L}). So far, we have witnessed only two strongly lensed and magnified \sneia, PS1-10afx and iPTF16geu \citep{2017Sci...356..291G}. We expect that the number will increase after these wide-field ground-based surveys come online \citep{2017ApJ...834L...5G}. These new strongly lensed \sneia at high-$z$ can be useful for similar studies as made here, before the launch of WFIRST.

\begin{acknowledgements}
We thank Ryan Chornock and Robert Quimby for providing useful information. R.A. and A.G. acknowledge support from the Swedish Research Council and the Swedish Space Board. The Oskar Klein Centre is funded by the Swedish Research Council. 
\end{acknowledgements}

\clearpage
\appendix

\end{document}